# A new approach to privacy-preserving clinical decision support systems

Thomas Attema[1,3,4,9], Emiliano Mancini[2], Gabriele Spini[3], Mark Abspoel[3,5],
Jan de Gier[1], Serge Fehr[3,4], Thijs Veugen[1,3], Maran van Heesch[1],
Daniël Worm[1], Andrea De Luca[8] Ronald Cramer[3,4], Peter M.A. Sloot[2,6,7]

1. Cyber Security & Robustness, TNO, The Hague, The Netherlands
2. Institute for Advanced Study, University of Amsterdam, Amsterdam, The Netherlands
3. Cryptology Group, CWI, Amsterdam, The Netherlands
4. Mathematical Institute, Leiden University, Leiden, The Netherlands
5. Philips Research, Eindhoven, The Netherlands
6. Complexity Institute, Nanyang Technological University, Singapore, Singapore
7. Advanced Computing, ITMO University, Saint Petersburg, Russia
8. Department of Medical Biotechnologies, University of Siena and Siena University Hospital, Siena, Italy
9. Corresponding author, thomas.attema@tno.nl

Abstract

*Background:* Clinical decision support systems (CDSS) are a category of health information technologies that can assist clinicians to choose optimal treatments. These support systems are based on clinical trials and expert knowledge; however, the amount of data available to these systems is limited. For this reason, CDSSs could be significantly improved by using the knowledge obtained by treating patients. This knowledge is mainly contained in patient records, whose usage is restricted due to privacy and confidentiality constraints.

*Methods:* A treatment effectiveness measure, containing valuable information for treatment prescription, was defined and a method to extract this measure from patient records was developed. This method uses an advanced cryptographic technology, known as secure Multiparty Computation (henceforth referred to as MPC), to preserve the privacy of the patient records and the confidentiality of the clinicians' decisions.

*Results:* Our solution enables to compute the effectiveness measure of a treatment based on patient records, while preserving privacy. Moreover, clinicians are not burdened with the computational and communication costs introduced by the privacy-preserving techniques that are used. Our system is able to compute the effectiveness of 100 treatments for a specific patient in less than 24 minutes, querying a database containing 20,000 patient records.

*Conclusion:* This paper presents a novel and efficient clinical decision support system, that harnesses the potential and insights acquired from treatment data, while preserving the privacy of patient records and the confidentiality of clinician decisions.

# 1 Background and significance

The constantly rising cost of national healthcare [1] associated to an aging population has highlighted the need for a revolution in traditional healthcare [2,3]. Most stakeholders (clinicians, healthcare providers, policy makers and patients) agree that the solution lies in new approaches in which technology and health information technology (HIT) play a critical role [4, 5]. HIT services aim to automate and optimize healthcare processes with the overall goal of providing a more effective treatment process for patients. One of the main hurdles of HIT is the need to preserve the privacy of the patients' data while using it to improve the quality of the tools that could be used by clinicians to provide better treatments. Legislation on the privacy of sensitive data, such as Health Insurance Portability and Accountability Act (HIPAA) and General Data Protection Regulation (EU) 2016/679 (GDPR), is becoming more and more stringent, affecting all parties who handle sensitive data. In order to accommodate the demands of the legislators without compromising the opportunities for healthcare offered by IT services, new solutions are needed to ensure patients' data is effectively processed while preserving their privacy. In this paper we focus on one specific category of HIT systems: *Clinical decision support systems* (CDSSs).

A CDSS is a system that provides clinicians, patients, and other individuals with intelligently processed disease-specific and patient-specific data. Several different categories of CDSSs can be found in literature, e.g. diagnostic tools, expert systems, workflow support, medication dosing support, order or billing facilitators, point-of-care alerts. The patient-specific data are presented in an appropriate form at appropriate times, in order to facilitate and improve treatment selection and quality of healthcare [6]. A CDSS is not meant to replace the clinician or make decisions in her or his place, but rather to help clinicians to determine the optimal treatment for each specific patient. Systematic reviews [7,8] reported that CDSSs significantly improved clinical practice. A review [7] on one hundred studies reported improvements for more than 62% of the trials on practitioner performance, reminder systems, drug-dosing systems and disease management systems. A review on seventy studies [8] reported a significant improvement of clinical practice in 68% of trials. Although general observations on the beneficial effects of CDSSs are limited by the heterogeneity of the studies, recent systematic reviews [9,10] report an improvement in health care processes in 148 randomized, controlled trials and in 85% of twenty-two studies respectively.

As a use-case to present our proposed solution to the problem of preserving the privacy of patients' data, we focus on an expert system for HIV treatment. The extreme variability in viral genotypes and the spreading of strains resistant to antiretroviral drugs make the prescription of optimal HIV-1 treatments a complex task; CDSSs are used in order to minimize or, ideally, prevent the prescription of suboptimal treatments. Some examples of relevant CDSSs range from simple quality improvement consultation programs like HIVQUAL-US [11] that monitors clinical performance, to more sophisticated data-driven systems like Euresist [12] and knowledge-based systems like the HIVdb Program [13]. In this paper we show the application of secure Multi Party Computation (MPC) to the "comparative Drug Ranking System" (cDRS). cDRS is a CDSS that helps to minimize the choice of sub-optimal HIV treatments by performing a meta-ranking analysis of three expert systems for HIV-1 genotypic drug resistance interpretation (ANRS, HIVdb, Rega) to resolve possible discordances between them [14-17]. The discordances in drug resistance between the three expert systems are not negligible [18,19], and are the result of different methodologies used by the systems to assess which set of mutations and mutational patterns lead to which level of drug resistance (susceptible, low-level resistance, intermediate resistance, high resistance) and of the limited amount of clinical data available for each specific set of mutations. A CDSS able to help clinicians in resolving such discordances is essential to avoid the administration of sub-optimal HIV treatments.

Research on the spread of the HIV epidemics has led to the development of tools (e.g., phylogenetic trees) able to correlate specific viral sequences in different patients and reconstruct with good accuracy the network of infections within a community [20]. In addition, transmission events between patients can be identified by analysing the viral genotypes, given the uniqueness of specific sets of mutations [21,22]. Hence, it is no wonder that strict privacy regulations prevent the sharing of patient data (e.g., viral genotype, treatments and their outcomes) that feed and improve these clinical decision support systems. Moreover, clinicians might not be able, or willing, to openly share their treatment decisions and their outcomes, even though such information might be beneficial for the decision-making process of their colleagues. In conclusion, there is a tremendous amount of valuable information that is unavailable to clinicians because of privacy and confidentiality constraints.

An ideal system should allow clinicians to compare their chosen treatment against the outcome of the treatments chosen by other clinicians for similar genotypes. Such a system should enable them to obtain actionable information that currently is not accessible at the scale and level of detail needed to maximize its usefulness. Such a system should also solve the issue of utilizing patient and clinicians' data in a secure way without leaking sensitive information.

The need for techniques to tackle this type of challenge is being addressed by public authorities as well: an important example in this sense is iDASH [23] (integrating data for analysis, anonymization, and sharing), a national center for biomedical computing established by the U.S. National Institutes of Health with the aim of developing algorithms and tools for the study of medical data in a privacy-preserving manner.

In this paper we present a solution that uses cryptographic techniques, namely a so-called *secure Multiparty Computation (MPC)* protocol, to achieve the desired functionality without violating any of the privacy constraints. Given $n$ mutually distrusting parties $P_1, \ldots, P_n$, each holding private inputs $x_1, \ldots, x_n$, the goal of MPC is to allow the parties to compute the value $f(x_1, \ldots, x_n)$ of a function f on their inputs, without revealing any other information than $f(x_1, \ldots, x_n)$, and without resorting to an external trusted party. Notice that these goals and requirements closely match the dilemma described above, as computing on medical data can provide very useful information, but privacy constraints forbid to freely exchange this data.

Early research in the 1980s [24-27] established the theoretical feasibility bounds for MPC; informally stated, this line of research proved that any function $f$ with finite domain and finite image can be evaluated securely in an MPC fashion. The precise security properties that can be achieved depend on the behavior of players and on the underlying communication model.

In the last one or two decades, research has shifted its focus to the efficiency and implementation of MPC. Since the first market-ready deployment of MPC in 2008 [28],

MPC solutions have been used in various practical contexts, e.g., stock market order matching [29], job market inquiries [30], and frequency bands auctions [31].

Application of MPC and other related cryptographic techniques in the medical domain has also been investigated in recent times, first in the study of general methods such as privacy-preserving data mining for joint data analysis between hospitals [32] and branching programs for privacy-preserving classification of medical ElectroCardioGram signals [33], then also in the presentation of specific use-case scenarios such as secure disclosure of patient data for disease surveillance [34], R-based healthcare statistics [35] and privacy-preserving genome-wide association study [36], privacy-preserving genome analysis [37] and search of similar patients in genomic data [38]. Moreover, various software suites and implementation frameworks for MPC have been made available [39-42].

Our MPC-based solution leverages private patient data and confidential clinicians' decisions, distributed amongst various hospitals and institutions, to provide additional information on the optimality of treatments. In particular, our solution allows clinicians to compare past treatments of 'similar' patients to find the optimal treatment for new patients preventing any unauthorized party, including the ones performing the computations, to access the input data.

## 2 Materials and Methods

### 2.1 Measuring treatment effectiveness from patient records

The viral genotype of a patient refers to the genetic sequence(s) of the HIV-1 virus strain that is most prevalent at the time of the blood test. The HIV-1 virus RNA genome contains 3 key regions that encode for enzymes critical to the life cycle of the virus: *protease (P), integrase (I)* and *reverse transcriptase (RT)*. Each region encodes for enzymes with 99, 288 and 560 amino-acids, respectively, all of which could in principle mutate. These mutations play an important role in the drug resistance of the virus strains.

Given an HIV-1 patient, the desired functionality will enable us to obtain treatment results of 'similar' patients, and therefore we need to define what it means for two patients, or two viral genotypes, to be similar, by defining a metric or distance function that quantifies this similarity. Since all expert systems indicate resistance to drugs on the basis of substitutions in the amino acid sequence of the wild type HIV-1, we need a way to compute the distance in the amino acid sequences of the viral proteins targeted by antiretroviral drugs: protease, reverse transcriptase and integrase. Metrics of distances between amino acid sequences are fairly complex and often assessed via neural networks [43]. The assignment of a suitable similarity metric is outside the scope of this paper and, for this reason, we have chosen to use a simplified viral genotype representation with a generic metric as a proof of concept. However, our solution is flexible, because it accommodates other representations and metrics. The main goal of this work is to demonstrate the applicability of MPC for HIV-1 CDSSs and open the door to many other applications that have been deemed impossible because of privacy and confidentiality restrictions.

From now on we shall represent viral genotypes $v$ as bit strings of a fixed length $N$, i.e., $v \in \{0,1\}^N$. We can think of each bit in this bit-string as an indicator for the presence or the absence of a specific mutation at a specific position.

Since there are only 97 relevant positions with commonly 1 or 2 resistance-associated substitutions [44] we can expect $N$ to be somewhere between 100 and 200. The distance between two viral genotypes $v_1$ and $v_2$ is defined by the Hamming distance between the bit strings,

$$H(v_1, v_2) = |\{i : v_1(i) \neq v_2(i)\}|,$$

which equals the number of positions at which the bits strings are different. Given this metric we can define two viral genotypes $v_1$ and $v_2$ to be similar if their Hamming distance is smaller than a certain threshold $B$, i.e., $H(v_1, v_2) < B$. Even though this metric is a simplification of the metrics used in practice, it is quite similar to the rule-based metrics used in the CDSSs of [14-17]. These CDSSs match viral genotypes based on the presence of resistance-associated substitutions in amino acid positions, which can be seen as a Boolean expression.

Suboptimal treatments of HIV-1 patients result in faster emergence of resistant strains and this emergence renders the treatment ineffective. Hence, a way to measure the effectiveness of a treatment $tr$ is by indicating the *time-to-treatment-failure* ($TTF_{tr}$). The $TTF_{tr}$ is defined as the time (in days) between the start of a therapy and either a therapy switch, a discontinuation of therapy or death [45,46]. Hence, given an HIV-1 patient with genotype $v$ we would, for example, like to compute the average $\overline{TTF_{tr}}(v)$ over all patients with similar genotype $v_i$, as an indication for the unknown true effectiveness measure $TTF_{tr}(v)$:

$$\overline{TTF_{tr}}(v) = \frac{1}{|\{i : H(v, v_i) < B\}|} \sum_{i : H(v, v_i) < B} TTF_{tr}(v_i).$$

Note that $TTF_{tr}(v)$ can only be computed for patients that have received a treatment $tr$ that has become ineffective and that the average $\overline{TTF_{tr}(v)}$ can only be computed when the set $\{i : H(v, v_i) < B\}$ is nonempty.

The described functionality could be realized by constructing a single database, containing a record for all completed treatments. Each record should contain the viral genotype, the administered treatment $tr$ and $TTF_{tr}$ — see Figure 1. From this database clinicians, should be able to query the average $TTF_{tr}$ for similar patients obtaining an indication for the effectiveness of that specific treatment $tr$.

However, these database records contain private information. The uniqueness of HIV virus mutations causes patients to be (almost) identifiable by only their viral genotype [21,22]. Moreover, patients with almost identical viral genotypes have most likely infected each other. Combined with other, possibly public sources of information, this data reveals a great deal of personal information. MPC offers a cryptographic solution to achieve the desired functionality without violating any of the privacy constraints.

| Record number: | 512 |
| --- | --- |
| Viral genotype: | 1000000111101… |
| Treatment ID: | 57 |
| TTF (in days): | 257 |

*Figure 1: Database containing sensitive information.*

## 2.2 Secure Multiparty Computation

Several considerations have to be made when applying MPC to a given problem. For instance, one may assume that parties $P_1, \ldots, P_n$ will behave semi-honestly (meaning that they may try to learn information on the other parties' inputs, but do follow the protocol), or that it is instead necessary to provide security against fully malicious players that deviate from the protocol instructions; one may focus on a two-party setting, or instead work with protocols that can support any number of parties. Another important parameter that varies among protocols is the number $t$ of corrupted parties that can be tolerated out of the total number $n$ of parties, with typical ranges being $t < n/3$, $t < n/2$ (honest majority), and $t < n$ ('unbounded' number of corrupted parties).

These are just some examples of the different considerations to be made. A remark of notable importance is that many desirable properties of MPC may negatively impact performance, or even be mutually exclusive, which means that the choice of an MPC protocol may be subject to important trade-offs. The reader can refer to [47] for a comprehensive discussion of MPC.

## 2.3 The MPC framework of our choice: SPDZ

We base our MPC solution on the SPDZ protocol [48,49]. The protocol is distinguished for its fast performance, and is implemented in a freely-accessible software suite [40,42] for UNIX-based systems[1].

SPDZ follows the so-called *share-compute-reveal* paradigm: each input $x_i$ of the function $f$ to be computed is 'dispersed' (or, formally speaking, *secret-shared*[2]) into $n$ pieces of data, called shares, each of which is assigned to a party; this process has the property that no information on $x_i$ can be extracted at all from a set of shares, unless such a set contains *all* shares (in which case $x_i$ can be completely recovered). Subsequently, parties execute a 'computation' protocol; as a result of this step, each party will have a share of the output $f(x_1, \ldots, x_n)$ of the function. Once all shares have been gathered, the output can then be reconstructed.

Other cryptographic techniques such as homomorphic encryption [50, 51] could potentially be of relevance for private data analysis, but we ruled out these alternatives, because of they would induce a huge computational overhead in our setting.

The share-compute-reveal approach is particularly well-suited for the client-server model we are interested in. Indeed, we do not wish to burden the input holders (i.e., the clinicians) with heavy computation, and would rather outsource this computation to external entities with a solid IT infrastructure. This can be readily achieved within the share-compute-reveal paradigm: the 'input' parties (clients) simply need to supply their secret-shared inputs to two or more 'computing' parties (servers), who will execute the computation protocol on these inputs, and then

---

[1] Notice that support for the SPDZ-2 software suite (implementing the eponymous MPC protocol) is being discontinued. Development moved to SCALE-MAMBA, another implementation of the SPDZ protocol.

[2] It is important to notice that 'sharing', here, is by no means a synonym of 'revealing'; on the contrary, it can be seen as a strong form of *encryption*.

communicate the shares of the output to the input parties, which can thus reconstruct the output.

It is important to remark that the SPDZ protocol does not, *per se,* distinguish between input and computing parties. A framework for MPC in a client-server model was presented in [52]; moreover, in [53] the SPDZ protocol was adjusted to the client-server setting.

The SPDZ protocol is divided into an 'offline' phase and an 'online' phase. The offline phase can be executed before the function inputs $x_1, ..., x_n$ are known, and its goal is to produce some secret-shared auxiliary data that will be used in the evaluation of $f$; producing this data can be a computationally-intensive process, but the since secret inputs are not required, this step can be executed during idle time and well before the actual secure computation will take place. Once the auxiliary data has been produced, the evaluation of $f$ can be performed very efficiently: this is of particular relevance for our use-case, where input parties (clinicians) need to obtain the output of the function $f$ within a matter of minutes, while preprocessing material can be produced in the background by the computing parties, so that the input parties are not burdened by this computational task.

## 3 Results

The functionality we have achieved utilizes HIV patient records to gain new insights in the effectiveness of HIV treatments. The MPC protocol ensures privacy of the patients and the confidentiality of the clinicians' treatment decisions.

The proposed solution distinguishes between 'input' parties, the clinicians supplying the database records, and 'computing' parties running the SPDZ protocol. The input parties additively secret-share their data records and distribute the shares amongst the computing parties (see Figure 2). As discussed in the previous section, this secret-sharing process has the property that private information can only be retrieved from the full set of shares; hence, if at least one of the $n$ computing parties does not collude with others, privacy is protected. We have chosen $n$ to be equal to 2, but for stronger privacy requirements we could choose to increase it.

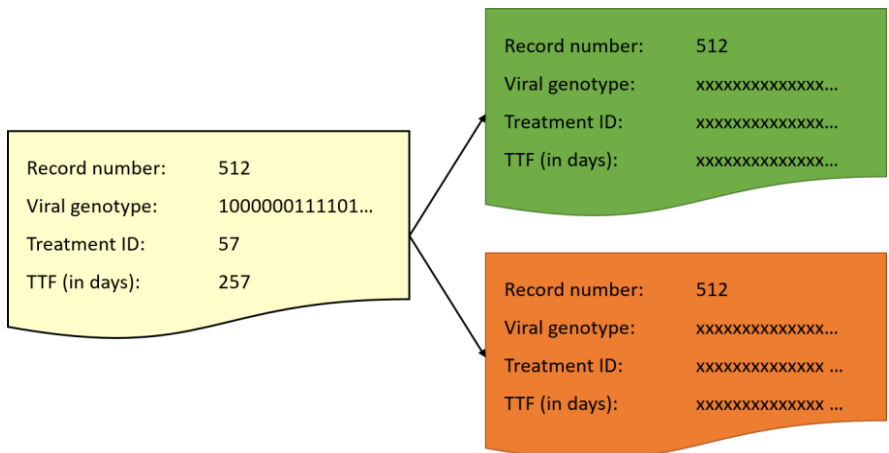

Figure 2: *Secret sharing database records.*

As a result, the two computing parties each hold a share of all the database records. SPDZ allows the evaluation of queries to this secret-shared database in such a way that only the output of the query (the average time-to-treatment-failure ($\overline{TTF}$) per treatment) is revealed to the clinician, and no additional information is leaked to either the querying clinician, or the computing parties. In order to protect the private information in the query (the viral genotype), we secret-share the query amongst the computing parties in a similar manner. The computing parties thus take as private inputs their shares of the database records and their share of the query. They do not reconstruct the result of the computation (the average $TTF$) themselves; instead, each of them sends their share of the result to the querying clinician who, in turn, recombines the shares to reconstruct the output. This way the result is only revealed to the clinician, and not to the computing parties (cf. Figure 3).

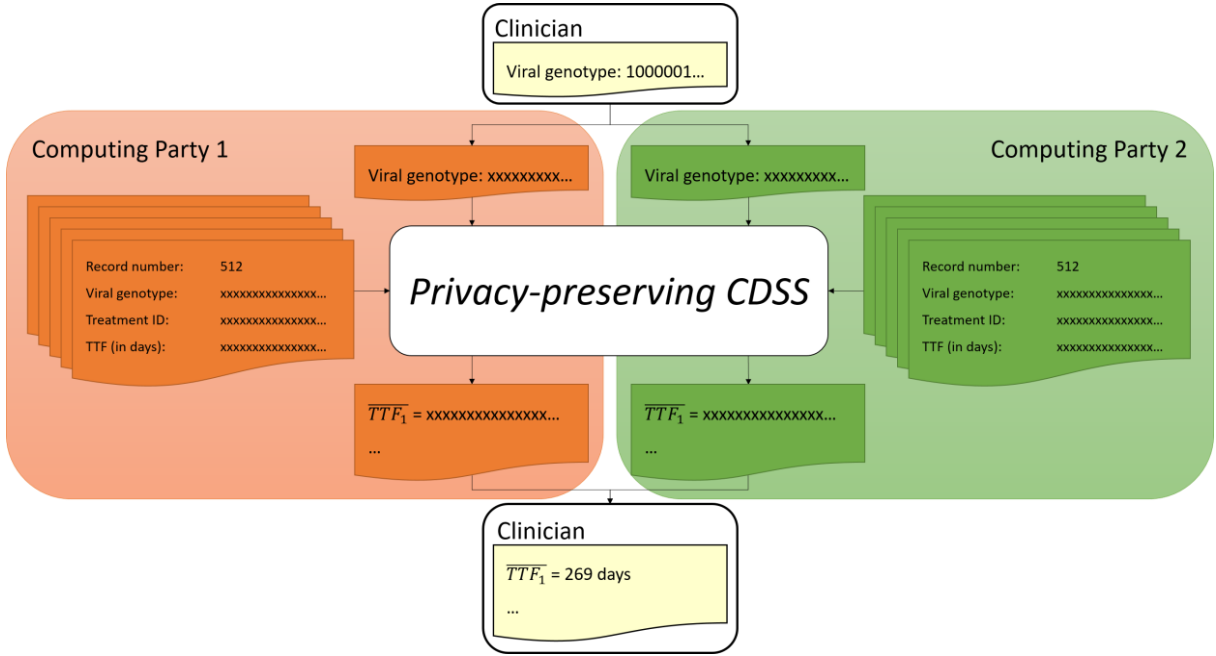

Figure 3: Query architecture of the privacy-preserving CDSS.

Our solution allows clinicians to compare their treatment of choice against the outcome of treatments previously chosen by other clinicians for patients with similar genotype, without revealing any private information to either the clinicians or the computing parties. In fact, the computing parties only learn the size and format of the database and the number of queries to the database. This system is secure as long as the two computing parties do not collude. We have implemented the online phase of the protocol by using the SPDZ software suite [40].

## 3.1 Performance – online phase

In comparison to implementing the functionality without privacy protection, using MPC introduces computational and communication overhead. The main reason for this unavoidable overhead is that, in an MPC protocol, the computation path should be oblivious, i.e. independent, of the input values, since it would otherwise leak information. Therefore conditional expressions such as if- and while-statements cannot be directly implemented: other (more expensive) techniques to obtain the same output in an oblivious way are required. Similarly, it might not be possible to implement efficient database searches that require ordering or re-structuring the database in a non-oblivious way.

We have evaluated the performance of the online phase of our protocol by deploying the computing parties on two different machines, each using one core of a i7-7567U CPU running at 3.50GHz and 32 GB of RAM, in a local network with 1 Gbit/s throughput. Moreover, we have instantiated the SPDZ protocol with 40-bit statistical security, 128-bit computational security and a 128-bit prime field.

The results in Figure 4 show the computation times that are needed for answering one query, for artificially-generated databases with sizes ranging from 100 to 20,000 records. The maximum 20,000 approximates the number of HIV-positive registered individuals in the Netherlands [54]; thus even though a single patient may give rise to more than one entry, we consider 20,000 to be an appropriate size to simulate a nation-wide database. The experiment is repeated multiple times, resulting in several data points per database size. Recall that per query we compute the average $TTF$ conditioned on 'similar' patients for 100 different treatments. Our current implementation can answer one query in less than 24 minutes if the database contains 20,000 patient records. The computational complexity scales linearly in the number of database records.

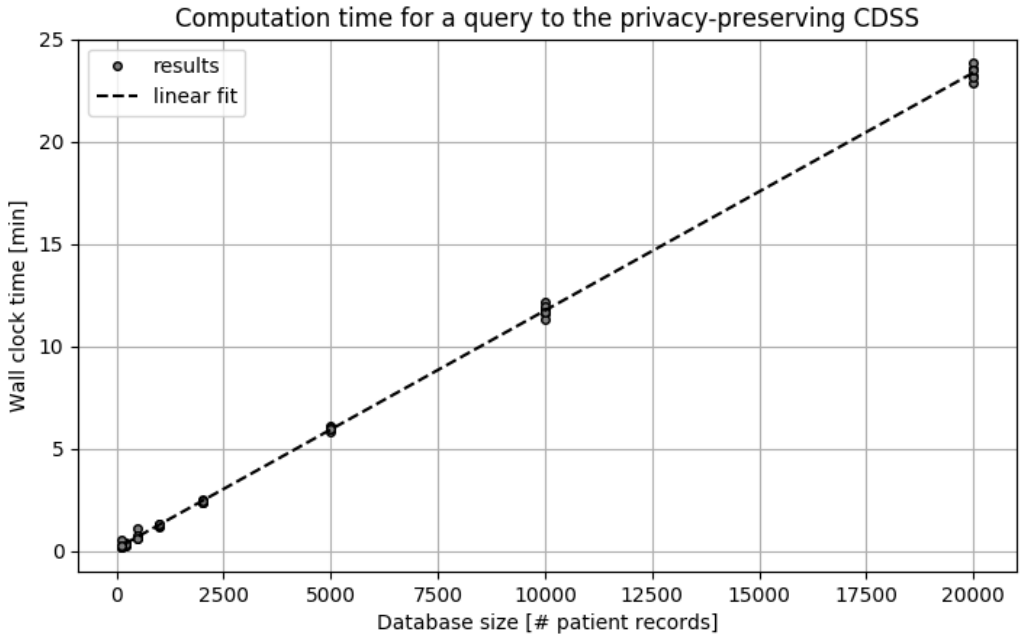

Figure 4: CDSS computation time.

## 3.2 Performance – offline phase

In the SPDZ protocol certain computational tasks are executed in the offline phase, that is independent of the MPC use-case and that can be implemented with existing protocols. For this reason, we have merely estimated the computational costs of it. The offline phase can be run at any time to generate a large database of preprocessed data which, in turn, is consumed during the online phase.

The performance of the offline phase can be quantified in the number of the so-called multiplication triples that are generated per second. In [55] various approaches for generating multiplication triples on i7-4790 and i7-3770S CPUs with 16 to 32 GB of RAM in a setting similar to ours (1 Gbit/s throughput) were evaluated. With two parties, 64-bit statistical security, 128-bit computational security and a 128-bit prime field, they generate 30,000 triples/s. To evaluate a single query on a database with 20,000 records approximately 40 million multiplication triples are required. In this setting these triples can thus be generated in approximately 22 minutes.

## 4  Discussion and Conclusions

We presented a novel approach for clinical decision support systems, making use of advanced cryptographic techniques to process private information without revealing it. Our solution allows clinicians to obtain valuable information on the several different HIV treatments that could be effective for a given patient, enabling them to query the success rates of treatments chosen by other clinicians for similar viral genotypes. By making use of MPC, we ensure both the privacy of the clinicians' treatment choices, thus releasing clinicians from liability concerns, and the privacy of patients, arguably making our solution in line with regulations on the use of medical information. Our solution can thus make valuable information finally available to clinicians, potentially improving HIV treatment and research on HIV drug effectiveness.

Towards a fully operational deployment, however, some points are yet to be addressed. Notably, the SPDZ software framework is designed for research purposes only, which means that our implementation should be audited and checked for vulnerabilities. For what concerns efficiency and scalability, we stress the fact that any CDSS for HIV treatment should produce a suggestion within minutes, since practitioners would typically query the system right after visiting a patient, and would expect an answer before the patient leaves their office. As shown in Figure 4, our solution answers a query within 24 minutes, for a database size roughly matching the number of HIV-positive registered individuals in the Netherlands [54]; while we consider this result to be sufficient for the proof-of-concept presented in this paper, some further work would be needed for a full-scale deployment. Notice that the running time of the implementation could be improved by several means, e.g. by using a low-level but very fast programming language such as C to encode the protocol, by further parallelizing the computation, or by making use of high-performance computing machines instead of consumer-level hardware. Another interesting point to be investigated is given by countermeasures against possible misuses of our solution, such as a malicious user that poses a high amount of queries in order to extract sensitive information from the received output.

## Acknowledgement

The authors would like to thank Pia Kempker for her valuable contributions to the early stages of this research. Moreover, the research activities that have led to this paper were partly funded by PPS-surcharge for Research and Innovation of the Dutch ministry of Economic Affairs and Climate Policy and partly funded by ERC Advanced Investigator Grant 740972 (ALGSTRONGCRYPTO).

and Applications of Cryptographic Techniques, Tel Aviv, Israel, April 29 - May 3, 2018 Proceedings, Part III, ser. Lecture Notes in Computer Science, Nielsen, J. B. and Rijmen, V., Eds., vol. 10822. Springer, 2018, pp. 158–189.

15